Conductivity scaling of the anomalous Hall effect in the altermagnetic semiconductor α-MnTe

Sara Bey[1], Shelby S. Fields[2], Nicholas G. Combs[2], Bence G. Márkus[1,3], Jiashu Wang[1], Liam Schmidt[4,5], Lincoln Curtis[4,5], Allecia Dodd-Noble[4,5], Alexander Poulin[4,5], Syed Mohammad Shahed[4,5], Resham Regmi[1,3], Mariia Holub[6,7], Phillipe Ohresser[6], Arun Bansil[4,5], Swastik Kar[4,5,8], Haile Ambaye[9], Valeria Lauter[9], László Forró[1,3], Cory D. Cress[10], Joseph C. Prestigiacomo[2], Nirmal Ghimire[1,3], Alberto de la Torre[4,5], Steven P. Bennett[2], Xinyu Liu[1], Badih A. Assaf[1]

[1] Department of Physics and Astronomy, University of Notre Dame, Notre Dame IN, 46556, USA.
[2] Materials Science and Technology Division, U.S. Naval Research Laboratory, Washington DC, 20375, USA
[3] Stavropoulos Center for Quantum Matter, University of Notre Dame, Notre Dame IN, 46556, USA.
[4] Department of Physics, Northeastern University, Boston, MA 02115, USA
[5] Quantum Materials and Sensing Institute, Northeastern University Innovation Campus, Burlington MA 01803, USA
[6] Synchrotron SOLEIL, L'Orme des Merisiers, 91190 Saint-Aubin, France
[7] Faculty of Physics and Applied Computer Science, AGH University of Krakow, 30-059 Kraków, Poland.
[8] Department of Chemical Engineering, Northeastern University, Boston 02115, USA
[9] Neutron Scattering Division, Neutron Sciences Directorate, Oak Ridge National Laboratory, Oak Ridge, TN, USA
[10] Electronics Science and Technology Division, U.S. Naval Research Laboratory, Washington DC, 20375, USA

**Abstract.** α-MnTe is a prototypical altermagnet exhibiting a strong anomalous Hall effect (AHE), despite having a nearly vanishing magnetization. Lately, sample-to-sample variations of the amplitude of the AHE have raised concerns of a possible defect related origin, especially in thin films. Here, we study the AHE in α-MnTe films grown on SrF$_2$ that have the crystal structure and m'm'm magnetic point group symmetry expected for bulk. By studying the scaling of the AHE with conductivity for those films and previously reported measurements in the literature, we find that sample-to-sample variations are well explained by a scaling law consistent with a hopping origin. Importantly, a comparison with other magnetic semiconductors reveals the colossal amplitude of the AHE of α-MnTe compared to its measured spontaneous magnetization from magnetometry and polarized neutron reflectivity. Our findings address the important fundamental question of the origin of the AHE of α-MnTe and further demonstrate the potential of altermagnets as promising spintronic materials.

Altermagnetism is a recently identified type of magnetic order. It arises in materials where two compensated magnetic sublattices cannot be related by spatial inversion or translation [1,2]. The electronic band structure of such materials supports a momentum dependent spin-splitting that is purely exchange induced. When spin-orbit coupling is included, altermagnets can allow a strong anomalous Hall and Nernst effect despite a vanishingly small magnetization [3–5]. Their properties, allowing spin generation, without magnetization and when spin-orbit coupling is small, have motivated device proposals spanning spintronics, magnonics and photonics [6] as well as fundamentally new quantum phenomena [7–13]. Semiconducting MnTe and metallic CrSb in their NiAs phase are the most notable altermagnets. Their exchange-induced band splitting has been observed and their magnetotransport properties have been studied in both single crystals and thin films [14–20].

Among these systems, α-MnTe is a widely studied semiconductor which realizes altermagnetism in its hexagonal α-phase. It was studied several decades ago and was reported to host an anomalous Hall effect, despite its nearly vanishing magnetic moment [21]. It has re-emerged as a promising spintronic material following reports of a tunable anisotropic magnetoresistance, a planar Hall effect, and most recently reports tying its strong anomalous Hall effect to its low magnetocrystalline symmetry [3,4,22–25]. In its magnetic ground state, Mn sublattices occupying different layers are coupled antiferromagnetically but cannot be mapped onto each other with simple translation or inversion making it an altermagnet. When the Neel vector $\vec{L} = \vec{M}_{Mn1} - \vec{M}_{Mn2}$ is along $<1\bar{1}00>$ [23,25–27], a spontaneous AHE is allowed by symmetry and has been observed in thin films and single crystals [3,4,28,29]. However, the magnitude, field, and temperature dependence of this AHE and its relation to the measured magnetization are all highly sample dependent [3,4,21,24,27–30]. It remains unclear whether these differences are intrinsic or defect driven as suggested in ref. [24,30]. Even among intrinsic mechanisms, a number of factors can influence the magnitude of the AHE, including but not limited to, variations in Fermi energy causing a changing Berry curvature, variations in the conductivity, variations in the magnetic order parameter and, to a lesser extent, variations in magnetic domain texture.

The anomalous Hall effect in systems that break time-reversal symmetry generally follows a scaling law that relates the Hall conductivity $\sigma_{xy}$ to powers of longitudinal conductivity $\sigma_{xx}$: $\sigma_{xy} = aM\sigma_{xx}^n$ [31]. The pre-factor here is a function of $\vec{M}$ the magnetic order parameter that generates the AHE. The physical meaning of $a$ depends on the mechanism governing the AHE. Establishing such a scaling law provides a powerful route to disentangle intrinsic mechanisms from disorder-induced contributions by comparing with the AHE of other known ferromagnets. In altermagnets, however, the spontaneous magnetization is virtually undetectable in films. It is thus impossible to disentangle whether differences in $\sigma_{xy}$ are caused by sample-to-sample variations in the magnetic order parameter or variations in the longitudinal conductivity, particularly when comparing thin films to single crystals.

Here, we investigate the origin of the AHE in α-MnTe thin films grown on SrF$_2$ (111) and we use a scaling analysis to elucidate its origin. The films grown on SrF$_2$ relax to a c/a ratio nearly identical to bulk (Fig. 1(a)). They also exhibit an X-ray magnetic circular dichroism (XMCD) exceeding what is expected from the measured magnetization consistent with the m'm'm magnetic point group which allows the AHE. We find that both the longitudinal and anomalous Hall conductivity of these α-MnTe films vary as a function of thickness, which allows us to construct a scaling plot. We combine our findings with the AHE reported in literature on single crystals to study the scaling law, and we find a scaling representative of a hopping-origin, expected for magnetic semiconductors for $\sigma_{xx} <10^4$ S/cm. The scaling then allows us to compare α-MnTe to other ferromagnetic semiconductors, revealing the colossal scale of its anomalous Hall conductivity $\sigma_{AHE}$ compared to its weak ferromagnetism. The correspondence between scaling in bulk and films that we reveal allows us to conclude that the AHE of α-MnTe is intrinsic and that sample-to-sample variations are due to variations in conductivity following a well-established scaling law.

Epitaxial α-MnTe films are grown by molecular beam epitaxy on SrF$_2$(111) substrates under Te rich conditions with a substrate temperature of 340°C - 390°C. X-ray diffraction measured about the (222) Bragg peak of the SrF$_2$ substrates confirms the formation of the α-phase through the observation of its (0004) Bragg peak (Fig. 1(b)). A reciprocal space map measured about the (513) SrF$_2$ and (11$\bar{2}$6) MnTe

Bragg reflections highlights that a 145nm α-MnTe film is not thermally-strained when grown on SrF$_2$(111) (Fig. 1(c)), unlike films grown on III-V wafers [26,32,33]. Table I lists the extracted a and c lattice parameters which yield a c/a ratio of 1.618, in agreement with single crystals. Additional structural characterization is published in a companion paper. The overall band structure of MnTe grown on SrF$_2$ also remains comparable to bulk [14–17] (see supplementary material) [34,35].

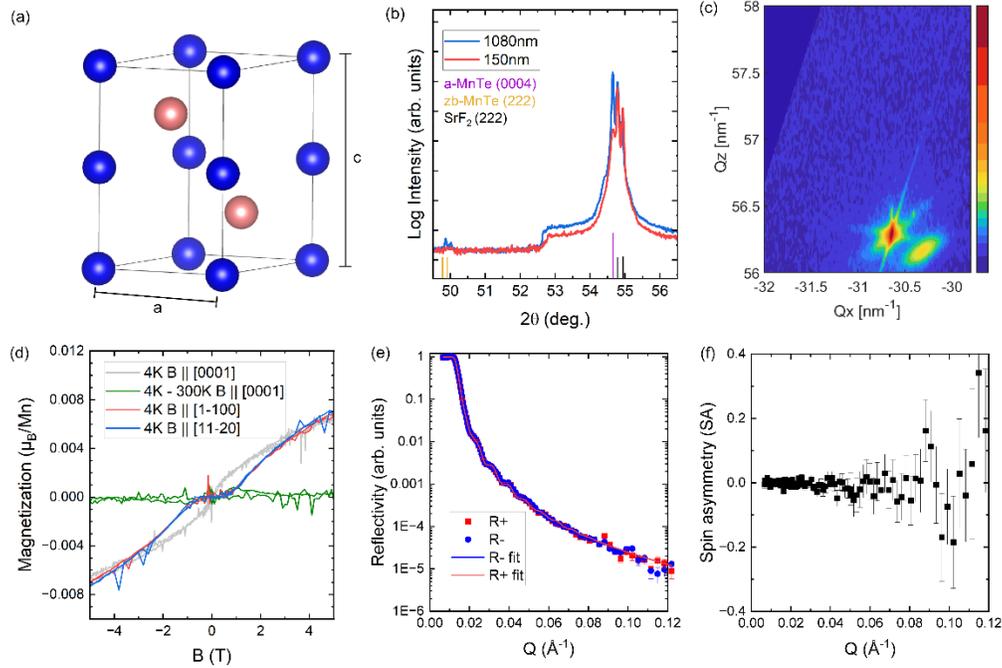

**FIG 1.** (a) Specular X-ray diffraction pattern near the SrF$_2$(222) peak for samples S1 (blue) and S2 (gray). The vertical lines below the plot indicate the expected peak positions for SrF$_2$, α-MnTe, and zincblende(zb)-MnTe. The Cu-K source is not monochromated. (b) Reciprocal space map of 145nm α-MnTe (S1) at 300K about the (११$\bar{2}$6) Bragg peak on MnTe. The RSM is acquired on a system equipped with a monochromated Cu-Kα source. (c) Magnetization of sample S2 measured at 4K with field applied along different directions, after subtracting a diamagnetic background (red/blue for in-plane and gray for out of-plane). Magnetization at 4K after subtracting raw signal at 300K (green line). (e) PNR reflectivity where R$_\pm$ refer to neutron spin parallel (+) or antiparallel (−) to the external field, respectively. (f) Polarized neutron reflectance spin asymmetry (SA) = (R$_+$ - R$_-$)/(R$_+$ + R$_-$) obtained from the experimental reflectivities shown in (e).

The magnetic properties of the MnTe films grown on SrF$_2$ (111) confirm that the magnetization nearly vanishes at zero field, as in previous work [23] (Fig. 1(d)). We carried out superconducting quantum interface device (SQUID) magnetometry measurements on the thick, 1080nm, α-MnTe epilayer S2 (see table I). Fig. 1(d) shows the magnetization versus magnetic field taken at 4K with field oriented in the ab-plane after subtracting the diamagnetic signal of the substrate. It is undetectable at low magnetic field (<10$^{-4}$μ$_B$/(u.c.)). A spin flop transition emerges at 0.8T when the field is applied in the ab-plane, but the net magnetic moment does not exceed 0.007μ$_B$/Mn even at 5T. Within our resolution, the magnetization behavior is identical for an in-plane field applied either parallel or perpendicular to one of the three equivalent easy axes ([1$\bar{1}$00] in Fig. 1(d)). A weak S-shaped field dependence is resolvable at 4K in the out-of-plane direction, but remains unchanged up to 300K (dashed line in Fig. 1(d)), suggesting that it is an extrinsic signal. Regardless, the out-of-plane magnetization is experimentally found to be smaller than 0.007$\mu_B$/Mn up to 5T. Importantly, the remanence in all directions is consistent with findings from single crystals, of a remanence on the order of $10^{-5}\mu_B$/Mn [3,28].

To further confirm that the magnetization is vanishingly small in thin films, we have carried out depth-sensitive polarized neutron reflectometry (PNR) at T=50K and B=1T on a Te-capped 77nm-MnTe/SrF$_2$2 film. We use neutrons with wavelengths spanning 0.2–0.8 nm and a high polarization of 98.5–99% [36]. Fig. 1(e) shows the raw reflectivity signal $R_+$ and $R_-$ for the neutron spin parallel (+) or antiparallel (−) to the external field and Fig. 1(f) shows the spin asymmetry ratio SA = ($R_+$ − $R_-$)/($R_+$ + $R_-$), measured as a function of the wave vector transfer Q . There is no resolvable splitting between $R_\pm$. The SA consistently overlaps with zero within error for Q as high 0.12Å$^{-1}$ despite significantly improved statistics and an extended Q-range compared to previous reports [24,37]. Without any analysis, our raw spin-asymmetry rules out the presence of a measurable bulk magnetization in α-MnTe/SrF$_2$ in the measurement range for which we have the best resolution. Fitting of the PNR is shown in the supplementary material [35].

A feature of altermagnets is their ability to host both XMCD and an AHE without a finite magnetization. Specifically, when $\vec{L}$ lies in the ab-plane of MnTe, a finite XMCD is only allowed if $\vec{L}||[1\bar{1}00]$, consistent with m'm'm magnetic point-group. We thus carry out x-ray magnetic circular dichroism (XMCD) at the Mn $L_{3,2}$ edge. Fig. 2(a) shows the zero-field x-ray absorption spectrum of α-MnTe grown on SrF$_2$ at $T$ = 4 K for light propagating along the (0001) direction. The zero-field XMCD response (Fig. 2 (b)) agrees well with that measured on α-MnTe/InP and reflects the altermagnetic contribution, [38] further confirming the absence of a detectable net out-of-plane moment. Similarly, the XMCD response (Fig. 2(b)) under a magnetic field B along the c axis remains consistent with reported values.

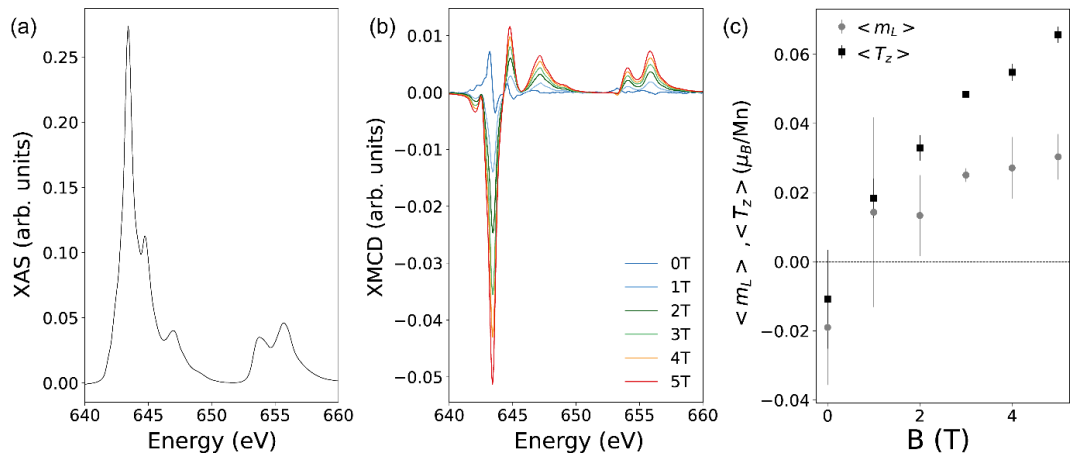

**FIG 2.** (a) Mn $L_{3,2}$ XAS measured on a capped α-MnTe film on SrF$_2$(111) at T = 4 K. (b) Normal incidence XMCD measured in total electron yield and (c) extracted <L$_z$> and <T$_z$> values as a function of applied magnetic field. The measurements are carried out a capped α-MnTe film grown on SrF$_2$.

To directly quantify the magnetic response, we apply XMCD sum rules to extract the Mn net magnetic moment [39]. We find $<S_z> \approx 0.16\ \mu_B/Mn$ at 5T, if we assume a negligible contribution of the anisotropic magnetic dipole (AMD), $<T_z> = 0$, consistent with MnTe grown on InP [40,41]. Notable however, the extracted value of $<S_z>$ is more than an order of magnitude larger than the magnetization at 5T m = 0.007 $\mu_B$/Mn measured by SQUID (see Fig. 1(d)). Hence, the magnitude of the in-field XMCD response far exceeds what can be expected from the field-induced canting of the Mn spins along the c-axis. Rather, recent theory work suggested that the XMCD response of altermagnets with zero net magnetization arises from finite $<T_z>$ which arises from the coupling between spin and quadrupolar orbital distributions [42,43]. To test this prediction, we re-evaluate $<T_z>$ from the XMCD sum rules by assuming $<S_z>(H) = (m(H) - <L_z>)/2$, with $<L_z>$ the extracted orbital angular moment. In Fig. 2(c) we show the dependence of $<T_z>$ and $<L_z>$ as a function of field. While at zero field $<T_z>, <L_z> \approx 0$, consistent with multiplet ligand field theory calculations [42], they both reach $<T_z> \approx 0.06\ \mu_B/Mn$ and $<L_z> \approx 0.03\ \mu_B/Mn$ at 5 T across the spin flop transition. We note that while $<T_z>$ behaves as an ordinary magnetic dipole, it carries no net magnetization. Our findings from XMCD confirm that MnTe grown on SrF$_2$ belong to the $m'm'm$ magnetic point group.

| Sample | Thickness (nm) | a (Å) – 300K | c (Å) – 300K | Reference |
|---|---|---|---|---|
| S1 | 145 | 4.146 | 6.713 | This work |
| S2 | 1080 | 4.15 | 6.714 | This work |
| S3 - Hall bar | 120 | | 6.711 | This work |
| on InP | 48 | 4.1708 | 6.686 | [4] |
| Bulk | 3 x 10$^5$ | 4.1486 | 6.7129 | [3] |

**Table I:** Lattice constants of α-MnTe grown on different substrates and bulk.

Magnetotransport measurements reveal a spontaneous AHE at low temperature, consistent with ref. [4,24] despite a nearly vanishing magnetization at zero magnetic field. Fig. 3(a) shows the anti-symmetrized Hall resistivity $\rho_{xy}(B)$ seen to be dominantly linear. Figures 3(b-d) plot the anomalous Hall resistivity $\rho_{AHE}$ extracted by subtracting a slope at high field from the Hall resistivity. Even in S2, where the upper bound on remanence cannot exceed 10$^{-4}\mu_B$/Mn (Fig. 1(d)), we measure a finite AHE at zero field, reaching 1μΩ.m (Fig. 3(c)). Qualitatively, all three samples exhibit an AHE at low temperature with a large coercive field. Both the coercive field and the spontaneous value $\rho_{AHE}$ at B=0 decay with increasing temperature. The decay is faster for the thickest sample S2, possibly due to its lower longitudinal conductivity. The reproducible AHE shown in Fig. 3(b-d) and the simultaneous measurement of a nearly vanishing magnetization in S2 (Fig. 1(c)), provide evidence that the AHE is not driven by the field dependence of a defect induced magnetization, but rather by that of the Neel vector $\vec{L}$. Our findings of both an AHE and an XMCD are consistent with the $m'm'm$ magnetic point group.

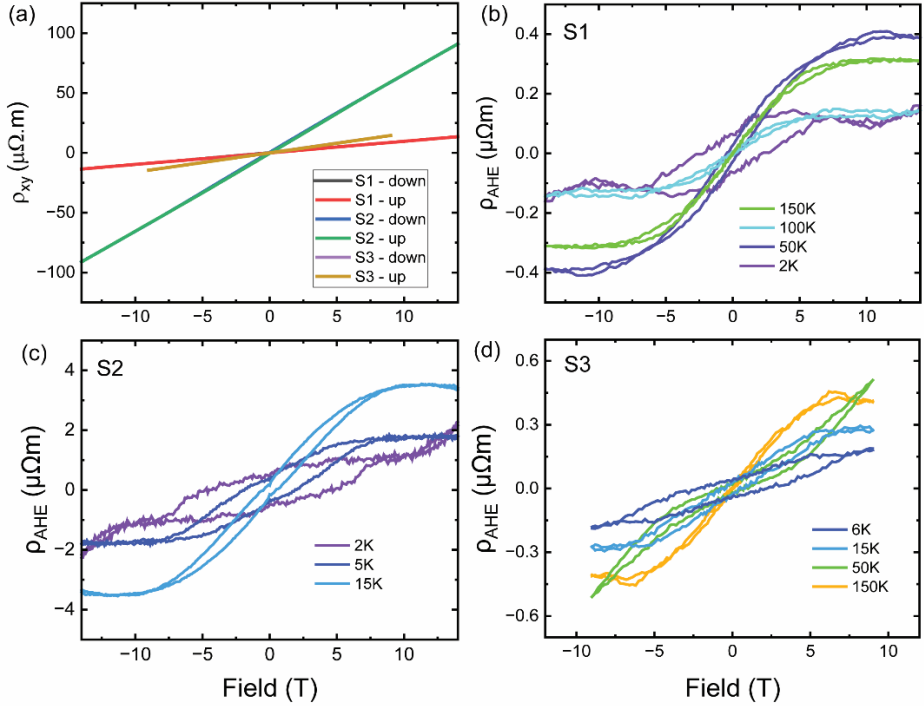

**FIG 3:** (a) Hall resistivity measured in samples S1-S3. (b-d) Anomalous Hall resistivity extracted after subtracting normal contribution for samples S1-S3. The temperature range considered varies in each case, because the AHE's temperature dependence is sample dependent.

The longitudinal resistivities measured in our films span an order of magnitude. Combined with measurements on single crystals from [3,28] and on thin films grown on InP(111) [4], they allow us to construct the scaling plot needed to reveal the underlying mechanism responsible for anomalous Hall effect in α-MnTe and its variation seen in the literature. A conductivity scaling relation has long served as a tool to diagnose the mechanism of AHE. In the insulating hopping regime ($\sigma_{xx} < 10^4$ S/cm), a universal scaling relationship of $\sigma_{AHE} \propto \sigma_{xx}^n$ with $1.33 \leq n \leq 1.76$ was developed for magnetic semiconductors [44]. The anomalous Hall effect is represented by hopping through triad clusters of impurity sites, which is the minimum number of sites to allow for an interference term representing broken time reversal symmetry. This interference is connected to the Berry phase [44]. In the metallic regime, $10^4 < \sigma_{xx} < 10^6$, a Berry curvature mechanism yields $\sigma_{AHE} \propto \sigma_{xx}^0$ [31]. We find a scaling relationship with $n = 1.39 \pm 0.08$ in MnTe (Fig. 4(a)), including our films and samples from refs. [3,4,28]. Such a scaling indicates that the hopping mechanism primarily drives the strength of the AHE [44]. It also quantitatively relates the disparities seen between bulk single crystals and thin films to the varying conductivity (see table II), without the need to invoke defect-induced ferromagnetism [24].

Crucially, the scaling plot also unveils the relation between $\sigma_{xy}$ and the order parameter, which we further use to shed light on whether the AHE in α-MnTe is defect-induced or intrinsic. We write down the magnetization and conductivity dependence of the AHE for ferromagnets, expressed logarithmically as:

$$\log(\sigma_{AHE}) = \log(aM) + n\log(\sigma_{xx})$$

Here $M$ is the magnetization (at zero field in this case) and $a$ is the proportionality constant between the AHE, the magnetization, and the longitudinal conductivity. In the intrinsic regime, it is determined by the net Berry curvature at the Fermi level [31]. In the hopping regime, two hopping paths give rise to an interference term for the transition rate that breaks TR symmetry. The interference term reflects the Berry phase accumulated over this hopping process [44].

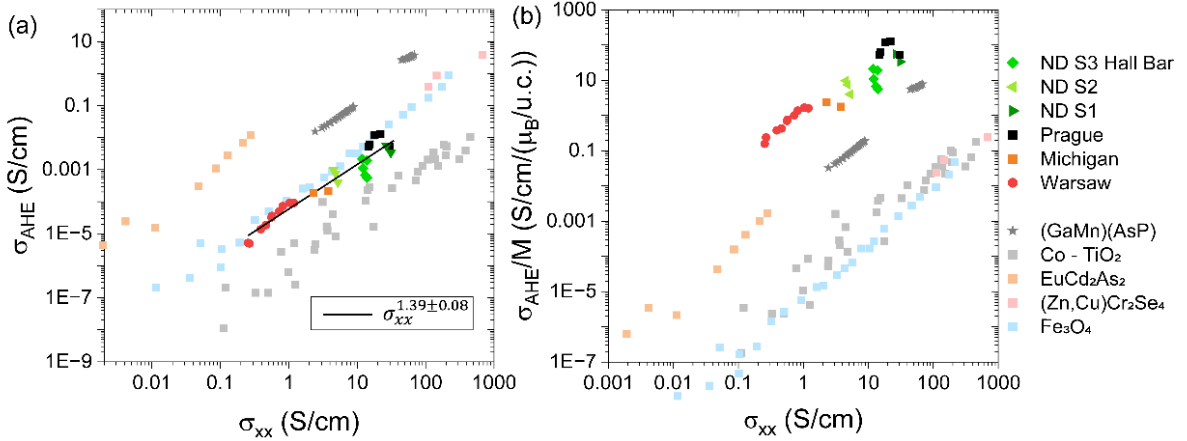

**FIG 4**. (a) Scaling of the anomalous Hall conductivity σ$_{AHE}$ and (b) σ$_{AHE}$/M and as a function of the longitudinal conductivity σ$_{xx}$. Data form this work is labeled ND. MnTe bulk (Warsaw and Michigan): [3,28]. MnTe on InP: (Prague) [4]). For films, we compute $\sigma_{AHE}/M$ by normalizing all data points (including MnTe on InP) by our detection limit (10$^{-4}$emu/u.c.) from the measurement shown in fig. 1(b). (Ga,Mn)(As,P): [45]. Rutile Co$_x$TiO$_2$: [46–50], we take $M = 0.06\mu_B/u.c.$ for all samples, as an estimate of the saturation magnetization with x=0.03. Fe$_3$O$_4$: [51], and $M = 19\mu_B/u.c.$ in [51]. (Zn,Cu)Cr$_2$S$_4$: scaling from [52]. Magnetization varies between 1 and $5\mu_B/(f.u.)$, points in (b) are normalized by equivalent of $2\mu_B/(f.u.)$ [53,54]. EuCd$_2$As$_2$: [55]. M=7μ$_B$/u.c. Among those materials only (Ga,Mn)(As,P) and MnTe exhibit a spontaneous anomalous Hall effect.

For comparison, Fig. 4 shows scaling data for other ferromagnetic semiconductors, EuCd$_2$As$_2$, Ga$_{0.94}$Mn$_{0.06}$(As,P), spinel (Cu,Zn)Cr$_2$Se$_4$, Fe$_3$O$_4$ and TiO$_2$ [45–55] in the hopping regime. We choose to include dilute magnetic semiconductors because they are systems where magnetic impurity doping has been shown to generate a strong anomalous Hall effect [45]. To remove the trivial dependence on magnetization and compare the AHE per unit magnetic moment, we replot the scaling as:

$$\log\left(\frac{\sigma_{AHE}}{M}\right) = \log(a) + n\log(\sigma_{xx}).$$

As seen in Fig. 4(b), α-MnTe exceeds all other semiconducting systems by two to four orders of magnitude. This establishes a *colossal anomalous Hall* response in α-MnTe, whereby its spontaneous transverse conductivity per unit magnetization is dramatically larger than in conventional ferromagnetic semiconductors. If one conservatively assumes that the AHE in α-MnTe is due to defect induced ferromagnetism, then it would require defects generating an anomalously large Berry phase [44] far beyond what is observed in established magnetic semiconductors - an implausible and highly unconventional scenario. The origin of the AHE in α-MnTe must then be dominantly intrinsic and related to the magnetocrystalline symmetry enabling altermagnetism. The symmetry allowed anomalous Hall conductivity of α-MnTe follows [56]:

$$\sigma_{AHE} \sim L_y(3L_x^2 - L_y^2)$$

Thus, our normalization by $M$ hides the fact that $M$ is a secondary order parameter since $\sigma_{xy}$ is in fact controlled by the direction of $\vec{L}$. We highlight here that although, $\vec{L}$ can be oriented along three equivalent $\{1\bar{1}00\}$ crystalline directions in different magnetic domains, they all yield an AHE of the same sign [56]. All our measurements are carried out after saturation to 12T, which is likely sufficient to cause domain imbalance between time-reversed domains and activate an AHE following the fourth-order exchange interaction argued to couple $\vec{L}$ and $\vec{M}$ in ref. [4,57].

| Sample | Temperature (K) | \|p\| (cm$^{-3}$) | μ (cm$^2$V$^{-1}$s$^{-1}$) | ρ$_{xx}$ (μΩm) | Ref. | M$_R$ (μ$_B$/u.c) |
|---|---|---|---|---|---|---|
| S1 | 2K | 6.44 x 10$^{18}$ | 27.4 | 353 | This work | Undetectable |
| S2 | 5K | 9.75 x 10$^{17}$ | 30.9 | 2067 | This work | <10$^{-4}$ |
| S3 | 6K | 3.9 x 10$^{18}$ | 21 | 762 | This work | Undetectable |
| on InP | 150 | 4.9 x 10$^{18}$ | 29 | 554 | [4] | Not reported. |
| Bulk | 237 | 4 x 10$^{17}$ | 8.6 | 17851 | [3] | 2.5×10$^{-5}$ |
| Bulk | 200 | -- | -- | 2644 | [28] | 6×10$^{-5}$ |

**Table II:** Properties from magnetotransport measurements of MnTe on different substrates and for bulk single crystals from [3,28]

In conclusion, magnetotransport and spectroscopy measurements on α-MnTe grown on SrF$_2$ (111), for which we are also able to detect the vanishing remanent magnetization, reveal that both the AHE and the XMCD of α-MnTe are anomalously large compared to its weak ferromagnetism. Our findings are consistent with the intrinsic origin related to altermagnetism [4] for both effects. A scaling analysis also shows that α-MnTe exhibits a colossal relative Hall response, exceeding that of established magnetic semiconductors by orders of magnitude. The conductivity scaling incorporating our thin films and single crystals places α-MnTe in the hopping regime of the AHE characteristic of low carrier density semiconductors [44] and accounts for the reported sample-to-sample variations. At elevated temperatures, although transport crosses over to multi-phonon assisted hopping, the scaling exponent remains unchanged since transverse response is controlled by symmetry and percolative network topology rather than by details of electron–phonon coupling. Overall, these results establish the AHE in α-MnTe as an intrinsic manifestation of altermagnetism and clarify its microscopic origin. α-MnTe thus provides a clear material platform in which large spontaneous transverse responses (AHE and XMCD) emerge from symmetry-protected altermagnetic order.

**Acknowledgements.** This material is based upon work supported by the Air Force Office of Scientific Research under award number FA9550-25-1-0319. This work was supported by the Office of Naval Research 6.1 Base Funding at the U.S. Naval Research Laboratory. XMCD experiments were performed on the DEIMOS beamline at SOLEIL Synchrotron, France (Proposal No. 20241735). S.K. and A.B. would like to acknowledge support from the National Science Foundation through the ExpandQISE award No. 2329067 and the Massachusetts Technology Collaborative through award number 22032. A portion of research used resources at the Spallation Neutron Source, a Department of Energy Office of Science User Facility operated by the Oak Ridge National Laboratory. The beamtime was allocated to instrument BL-4A

MAGREF on proposal number IPTS-32732.1. This manuscript has been authored by UT-Battelle, LLC under Contract No. DE-AC05-00OR22725 with the U.S. Department of Energy. The United States Government retains and the publisher, by accepting the article for publication, acknowledges that the United States Government retains a non-exclusive, paid-up, irrevocable, world-wide license to publish or reproduce the published form of this manuscript, or allow others to do so, for United States Government purposes. The Department of Energy will provide public access to these results of federally sponsored research in accordance with the DOE Public Access Plan (http://energy.gov/downloads/doe-public-access-plan).